# Exploring the Effectiveness of Deep Features from Domain-Specific Foundation Models in Retinal Image Synthesis

Zuzanna Skórniewska[1][0009-0000-9050-383X] and Bartłomiej W. Papież[1][0000-0002-8432-2511]

[1] Big Data Institute, Nuffield Department of Population Health, University of Oxford
zuzanna.skorniewska@ndph.ox.ac.uk

**Abstract.** The adoption of neural network models in medical imaging has been constrained by strict privacy regulations, limited data availability, high acquisition costs, and demographic biases. Deep generative models offer a promising solution by generating synthetic data that bypasses privacy concerns and addresses fairness by producing samples for under-represented groups. However, unlike natural images, medical imaging requires validation not only for fidelity (e.g., Fréchet Inception Score) but also for morphological and clinical accuracy. This is particularly true for colour fundus retinal imaging, which requires precise replication of the retinal vascular network, including vessel topology, continuity, and thickness. In this study, we investigated whether a distance-based loss function based on deep activation layers of a large foundational model trained on large corpus of domain data, colour fundus imaging, offers advantages over a perceptual loss and edge-detection based loss functions. Our extensive validation pipeline, based on both domain-free and domain specific tasks, suggests that domain-specific deep features do not improve autoencoder image generation. Conversely, our findings highlight the effectiveness of conventional edge detection filters in improving the sharpness of vascular structures in synthetic samples.

**Keywords:** Colour fundus imaging, VQ-GAN, Generative Modelling.

## 1 Introduction

Colour fundus imaging is a non-invasive, affordable imaging technique routinely used in ophthalmologic clinics for retinal diagnostics. However, this imaging shows promise beside clinical practice - in biomarker discovery, with discoverable features associated with cardiovascular [1], [2] neural [3], [4] and ophthalmic [3], [5] conditions. While the use of deep learning for biomarker extraction would remove the need for manually crafted features, there have only been a few works demonstrating a fully automated deep learning approach [1], [3], [6]. The main barrier to further progress is the limited availability of large-scale datasets, as most retinal datasets, such as DRIVE [7] and STARE [8], are small fully labelled sets primarily used for segmentation tasks.

The scarcity of data can potentially be addressed by generating synthetic samples using generative neural networks. This approach may also help mitigate privacy concerns and reduce sampling biases stemming from a lack of diversity in existing clinical studies [9]. While generative neural networks, such as VQ-GAN [10] or more recently diffusion models [11], [12], [13], have proven capable of producing high resolution realistic and diverse natural images, as validated by e.g., Fréchet Inception Score, there is currently no standardized method for assessing the morphological accuracy of medical images, including synthetic colour fundus images. This includes validating aspects such as vessel continuity, differentiation between arteries and veins, and clinically relevant vascular thickness and topology.

To date, the use of generative models for medical imaging have been predominantly synthetic brain [14], [15], [16] and breast MRI [17], tissue [18], cell [19], and cancer [20], [21] imaging data. The image generation is controlled and validated for its semantic correctness associated with the domain in question via incorporation of domain-specific supervision losses [16], meta-data conditioning [16], [17], [19], self-supervision strategies [15], [18] and downstream validation on external clinical data or extracted biomarkers [14], [17]. However, colour fundus image synthesis has to date been proportionally unaddressed, with a few notable exceptions [22], [23], [24].

This paper explores colour fundus generation using an encoder-decoder model, inspired by VQ-GAN architecture [10] capable of generating high-quality synthetic colour fundus images and trained with a range of auxiliary loss functions, which enforce either perceptual fidelity, vessel continuity or domain-specific morphological & clinical feature preservation. We extend the standard validation beyond fidelity metrics to a range of relevant downstream clinical prediction tasks, as well as a direct comparison of retinal morphological features relating to vascular networks and the optic disc.

Specifically, inspired by the success of training guidance based on deep features from networks trained on large natural image datasets to enforce perceptual similarity, coined as perceptual loss [25], this paper introduces a novel loss term, conceptually similar to perceptual loss. The novel loss term leverages a RETFound [3] backbone, a large vision transformer trained via self-supervised learning on a dataset of 1.6 million unlabelled colour fundus images. We refer to this as *RETFound loss*, which directly optimizes the imaging latent space. By minimizing the distance between synthetic images and real counterparts in the latent space, we hypothesise that the generated images retain deep features associated with domain-specific imaging fidelity – relating to retinal morphology and clinical information contained within the image. Our hypothesis is based on the strong performance of the backbone model – RETFound in various downstream classification tasks on external clinical metadata, supporting its effectiveness in this context.

## 2 Background

### 2.1 VQ-GAN

Our model follows the architecture of VQ-GAN encoder-decoder model [10] with latent space quantization and supervision by a CNN-based patched discriminator, trained concurrently as the autoencoder in an adversarial way. The encoder and decoder are convolutional, featuring down-sampling and up-sampling layers respectively, followed by ResNet [26] blocks. Additionally, an attention block is applied after down-sampling in the encoder and before up-sampling in the decoder. Specifically, the model comprises of an encoder, which projects an input image, $x \in R^{H \times W \times 3}$, to a compressed latent space, $z \in R^{h \times w \times n_z}$, where $n_z$ is the dimensionality of the embedding space and $h = \frac{H}{c}$ and $w = \frac{W}{c}$, where H, W are an image height and width, and c is the compression magnitude based on encoder's depth and the input image resolution. This latent representation contains $h \times w$ vectors $z_{i,j} \in R^{n_z}$. In the quantization stage, each of $z_{i,j}$ is mapped to its nearest neighbour, $\hat{z}_j \in R^{n_{\hat{z}}}$, based on the Euclidean distance, where $\hat{z}_j$ is chosen from the set of K codebook entries. The codebook $Z = \{\mathbf{e_k} \in R^{n_z}\}_{k=1}^{K}$ is a discrete set of learnable embedding vectors used to quantize the continuous latent representations; it is optimized jointly with the encoder and decoder by minimizing a loss encouraging quantized vectors to stay close to the encoder output. Upon the quantization stage, the parametrised embedding space is fed through a decoder, which reconstruct the information in the compressed latent space back to pixel space.

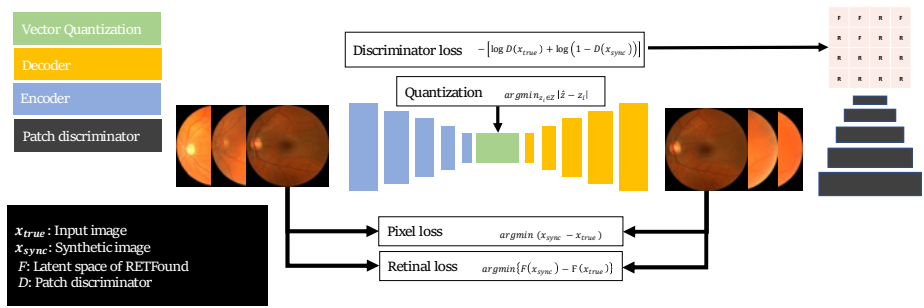

**Fig. 1.** The model is a standard VQ-GAN with an autoencoder-decoder architecture, incorporating quantization reparameterization in the latent space and a CNN-based patch-based discriminator supervision. The proposed training strategy introduces a loss function defined in the latent space of RETFound [3] trained on 1.6 million retinal images.

This encoder-decoder model, referred to as the *generator*, is trained using GAN loss equilibrium, with a CNN patch-based discriminator enforcing the generation of realistic-looking samples. The generator's training process is driven by gradient updates from three sources: the pixel loss (Hinge loss), the GAN penalty term, and the novel RETFound loss term. The RETFound loss minimizes the distance between embedding vectors of synthetic and real samples in the RETFound latent space, ensuring that both images capture similar semantic information related to the retina. In parallel with training the generator and discriminator, the set of codebook entries is optimized on the training data via *commitment loss [27]*.

### 2.2 RETFound

The RETFound model, similar to VQ-GAN, follows an encoder-decoder architecture. The encoder is a large vision transformer [28] (ViT-large) with 24 transformer blocks, which partitions the input image into 16×16 patches and encodes it into a 1024-dimensional embedding vector. The decoder is a small vision transformer (ViT-small) with 8 transformer blocks. The model was trained on a curated dataset of 1.6 million unlabelled colour fundus retinal images with a resolution of 224. The authors compare the model performance with models trained using a range of self-supervised strategies, concluding the mask autoencoder as the one yielding the best performance.

### 2.3 Perceptual loss

Perceptual loss was introduced as a stronger alternative to model-free perceptual distance metrics, such as the Structural Similarity Index (SSIM), which struggle to accurately capture certain types of image degradation, like blurring [25]. Instead of relying on traditional similarity measures, perceptual loss quantifies perceptual similarity between image pairs using the L1 distance between deep feature activations from large CNN models trained on natural images. This approach is based on the idea that deep network layers, having learned to recognize meaningful patterns in images, are inherently sensitive to degradations that align with human perception.

## 3 Methods

### 3.1 VQ-GAN implementation details

Our model was trained on 90,344 left and right colour fundus images at a resolution of 256×256 from the UK Biobank. The encoder compresses the 3-channel input images using scaling factors of [1,2,2,4], with each down sampling followed by one ResNet block, progressively reducing image's spatial dimensions to 32×32. This results in a latent representation consisting of 1,024 embedding vectors, each with 256 channels. During the quantization stage, the embedding channels are reduced from 256 to 128, and each of the 1,024 vectors is mapped to one of 512 learned codebook entries. This represents a mild compression, which retains essential information while eliminating redundancy present in the highly correlated pixel space, ensuring that the latent representation captures the most meaningful features of the image.

### 3.2 Retinal feature extraction

Assessment of morphological correctness and clinical usefulness of synthetic colour fundus imaging is based on retinal features extracted by an external end-to-end model, AutoMorph [6], which extracts a set of 45 interpretable features, relating to geometric properties of the optic disc and cup (height and width) and the retinal vascular network (e.g., vascular thickness or fractal dimension).

### 3.3 RETFound loss

Our *RETFound* loss is based on the RETFound model trained on Moorfields Hospital data using masking training strategy. The loss is computed as the L1 distance in the 1024-dimensional embedding space – the activation layer of the encoder's last transformer block. Since the RETFound model was trained on images with a resolution of 224, our images are resized accordingly.

### 3.4 Edge loss

Most handcrafted retinal biomarkers are linked to retinal vascular networks, as evidenced by tools like AutoMorph, where the majority of features are vascular-related. Inspired by this, we implemented an L2-distance loss function between the "vesselness maps" of an image pair. Specifically, the vesselness response is computed on the green channel using the Meijering filter [29]— an intensity-based filter based on the eigenvalues of the Hessian matrix of the image gradient. This filter response effectively highlights tube-like structures, such as blood vessels.

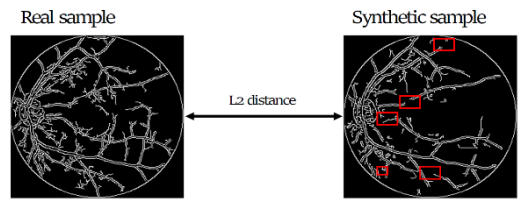

**Fig. 2.** Meijering filter-based edge loss: Computes the L2 distance between vesselness scores to enforce topological correctness and vessel continuity. Note that the visualization shows a thresholded signal, while the loss operates on the continuous vesselness response.

### 3.5 Perceptual loss

Perceptual loss is computed using a VGG-16 [30] backbone trained on ImageNet for classification. The loss is calculated as the sum of L2 distances between feature maps extracted from the 4th, 9th, 16th, 23rd, and 30th activation layers of the network.

## 4 Experiments

We performed a comprehensive validation of the generated samples, beginning with standard validation methods using commonly used fidelity metrics, followed by evaluation on downstream classification tasks incorporating additional clinical metadata, and finally, direct comparison of extracted biomarkers related to retinal vasculature and optic disc properties.

### 4.1 Image Fidelity Evaluation

To validate the quality of synthetic images, we used several metrics: Fréchet Inception Distance (FID), Maximum Mean Discrepancy (MMD), and Multi-scale Structural Similarity Index Measure (MS-SSIM). These metrics respectively evaluate the fidelity of synthetic samples, their coverage relative to real samples, and their structural similarity. Each metric is based on 20,000 synthetic and real samples. FID is based on the latent space of InceptionV3 [31] trained on natural images (ImageNet), while MMD is based on Gaussian kernel operating in pixel space.

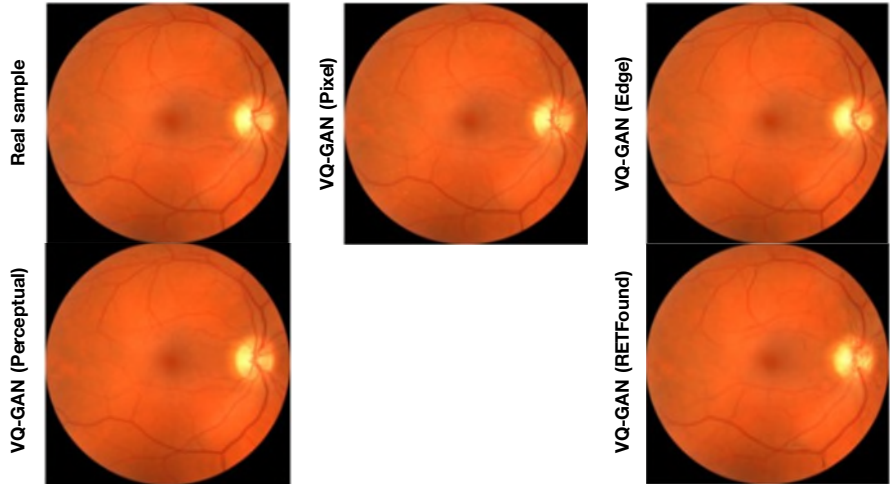

**Fig. 3.** Qualitative comparison of tested VQ-GAN models on a single sample. All models achieve good reconstruction, but VQ-GAN (Edge) and VQ-GAN (Perceptual) exhibit slightly sharper contrast, enhancing vessel detection, whilst VQ-GAN (RETFound) struggles to replicate the optic cup.

As shown in Table 1, VQ-GAN trained with perceptual loss achieves the FID closest to that of real samples, followed closely by VQ-GAN trained with edge loss. Interestingly, incorporating deep features from the RETFound model led to only a slight improvement in the fidelity of synthetic samples compared to the pixel-only loss baseline. Unlike the FID score, both MMD and MS-SSIM offered limited insight, as they produced similar values across all four models. Specifically, MMD remained consistently low (below 0.05), indicating satisfactory sample coverage, while MS-SSIM values were higher than those of real samples. This increase in structural similarity may be attributed to the filtering effect of autoencoders, which generate synthetic images that are free of imaging artifacts found in real data.

**Table 1.** Commonly used metrics to assess synthetic samples fidelity and diversity. VQGAN (RETFound) - model trained with RETFound loss; VQGAN (Perceptual) - trained with perceptual loss; VQ-GAN (Pixel) - trained only with pixel-based losses; VQ-GAN (Edge) - trained with edge loss. MS-SSIM is reported as mean [standard deviation].

| Model | FID ↓ | MMD ↓ | MS-SSIM ↑ |
|---|---|---|---|
| Real | 0.64 | 0.022 | 0.706 [0.065] |
| VQ-GAN (RETFound) | 11.43 | 0.020 | 0.968 [0.010] |
| VQ-GAN (Perceptual) | **2.45** | 0.019 | 0.975 [0.007] |
| VQ-GAN (Pixel) | 19.95 | 0.020 | **0.980 [0.006]** |
| VQ-GAN (Edge) | 5.47 | **0.018** | 0.975 [0.007] |

## 4.2 Evaluation on downstream tasks

Beyond assessing visual fidelity, synthetic medical images must also preserve clinical relevance and morphological accuracy. To validate this, we quantified the predictive power of retinal features extracted using AutoMorph on various cardiovascular risk factors, specifically age, sex, BMI, blood pressure, and blood glucose levels, which were previously shown to be directly predictable from colour fundus images [1]. The evaluation follows the standard *Train-Synthetic, Test-Real* framework [32]. As shown in Table 2, all models except VQ-GAN (RETFound) achieved the lowest error in two variables each. VQ-GAN (Perceptual) had the best results for age and BMI, VQ-GAN (Edge) for systolic blood pressure and HbA1c and the baseline VQ-GAN (Pixel) for diastolic blood pressure and glucose levels. Notably, VQ-GAN trained with RETFound deep features underperformed, yielding worse results than even the baseline VQ-GAN (Pixel). However, none of the synthetic models achieved error rates as low as those based on real data, highlighting a persistent and yet unresolved generative inconsistency.

**Table 2.** Validation of synthetic samples on prediction tasks on external data, reported as (MAE - mean (standard deviation) for age, BMI, diastolic/systolic blood pressure, HbA1c and glucose, and F1-score for sex.

| Validation variable | Real | VQ-GAN (RETFound) | VQ-GAN (Perceptual) | VQ-GAN (Pixel) | VQ-GAN (Edge) |
|---|---|---|---|---|---|
| Sex | 0.52 (0.01) | 0.53 (0.01) | 0.51 (0.02) | 0.51 (0.01) | 0.51 (0.012) |
| Age | 7.72 (0.13) | 9.32 (1.68) | **7.60 (0.27)** | 8.00 (0.43) | 7.94 (0.18) |
| BMI | 4.01 (0.06) | 5.21 (0.68) | **4.29 (0.27)** | 4.31 (0.24) | 4.30 (0.33) |
| Diastolic BP [mmHg] | 9.12 (0.18) | 11.84 (0.80) | 10.17 (0.47) | **9.60 (0.29)** | 10.58 (0.93) |
| Systolic BP [mmHg] | 16.98 (0.22) | 20.88 (4.26) | 18.52 (0.71) | 18.01 (0.83) | **17.55 (1.71)** |
| HbA1c [mmol/mol] | 4.27 (0.09) | 6.93 (1.34) | 5.28 (0.57) | 4.98 (0.77) | **4.59 (0.25)** |
| Glucose [mmol/L] | 0.63 (0.06) | 1.48 (0.79) | 0.77 (0.16) | **0.74 (0.09)** | 0.78 (0.12) |

## 4.3 Morphological Evaluation

In addition to evaluating extracted retinal features for predicting external clinical variables, we also directly assessed the similarity of sample distributions. Specifically, for a set of 45 AutoMorph features, we computed the mean and standard deviation for each sample. For non-real samples, we further assessed the probability of belonging to the same distribution as real data using a permutation test, where a high p-value indicates a faithful recreation of the real sample distribution. This analysis was conducted on a set of 5,000 randomly selected images. Figures 4-6 present the results for all variables, with features categorized as follows: optic disc and cup features (Figure 4; Table S1), vascular width (Figure 5; Table S2), vessel density (measured as the proportion of vessel to non-vessel pixels; Figure 6 – right top panel; Table S3), fractal dimension (expressing vascular complexity; Figure 6 – left top panel; Table S4), and vascular calibre metrics, including CRAE (central retinal arteriolar equivalent), CRVE (central retinal venular equivalent), and AVR (arteriolar–venular ratio) (Figure 6 – bottom panel; Table S5). Figures 4–6 show the distribution of synthetic features relative to the real sample distribution (z-score normalized). Coloured distributions indicate substantial overlap with the real data, as determined by a permutation test (p-value > 0.05). Exact summary statistics are provided in the Supplement.

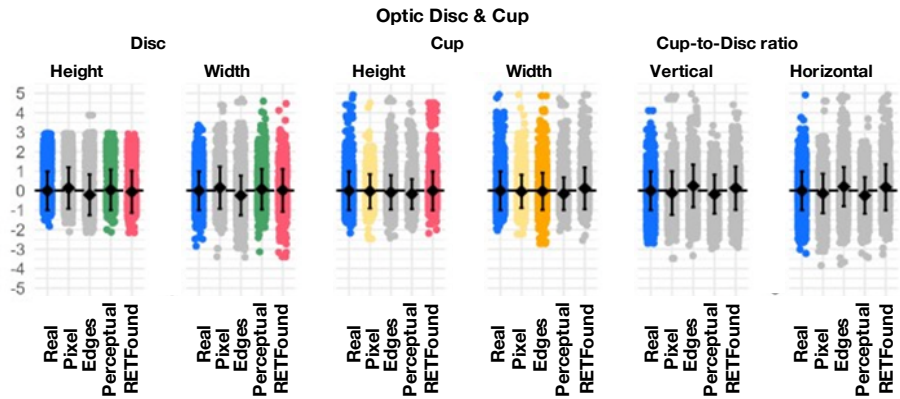

**Fig. 4.** Comparison of retinal features related to the optic disc and cup. Each boxplot shows the mean and standard deviation for each model. Non-grey distributions indicate samples' overlap with the real data distribution, as supported by a permutation test (p-value > 0.05). All features are z-score normalized relative to the real sample.

As indicated by Figures 4-6, VQ-GAN trained with perceptual loss demonstrated the highest consistency, yielding 5 high p-values, matching only a fraction of all 45 features. This was followed by VQ-GAN (Edge) with 4 high p-values, VQ-GAN (Pixel) with 3, and VQ-GAN (RETFound) with 3. These findings align with the models' performance in downstream prediction tasks and further highlight the underperformance of fundus image generation when leveraging deep features from a foundational model trained on target domain data.

Notably, VQ-GAN (Edge) achieved a relatively high number of matches compared to VQ-GAN (Perceptual), while requiring significantly lower computational resources and being more data-efficient. Since edge detection is a weight-free process that does not require extensive training on large datasets, it offers a practical alternative to perceptual loss.

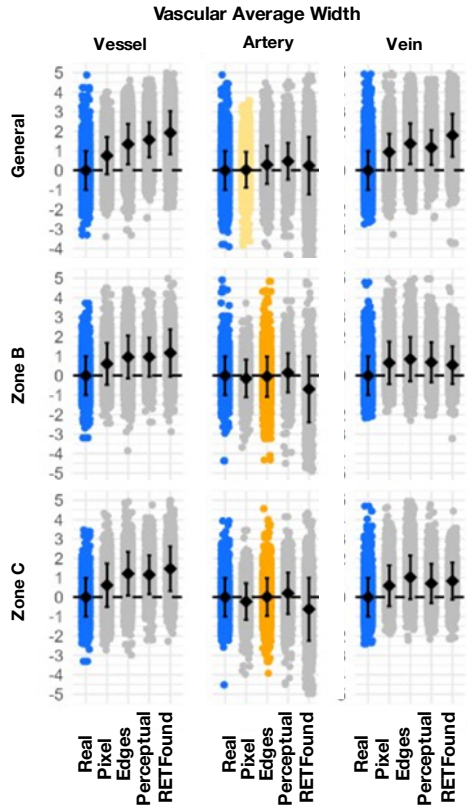

**Fig. 5.** Comparison of retinal features related to vascular average width. Each boxplot shows the mean and standard deviation for each model. Non-grey distributions indicate samples' overlap with the real data distribution, as supported by a permutation test (p-value > 0.05). All features are z-score normalized relative to the real sample.

Interestingly, synthetic data performance varied considerably across different retinal features. Features related to the optic disc and cup were well represented, whereas those linked to average vascular width and fractal dimension were consistently less accurate. Specifically, synthetic models tended to generate vasculature that appeared thicker than in real samples, as shown in Figure 5. While VQ-GAN (Edge) performed best in this category, it still only matched two out of nine features. Conversely, synthetic samples consistently produced lower values for fractal dimension features, suggesting a tendency to simplify vascular complexity compared to real data, as shown in Figure 6, with VQ-GAN (RETFound) having the largest deviation from the real sample distribution.

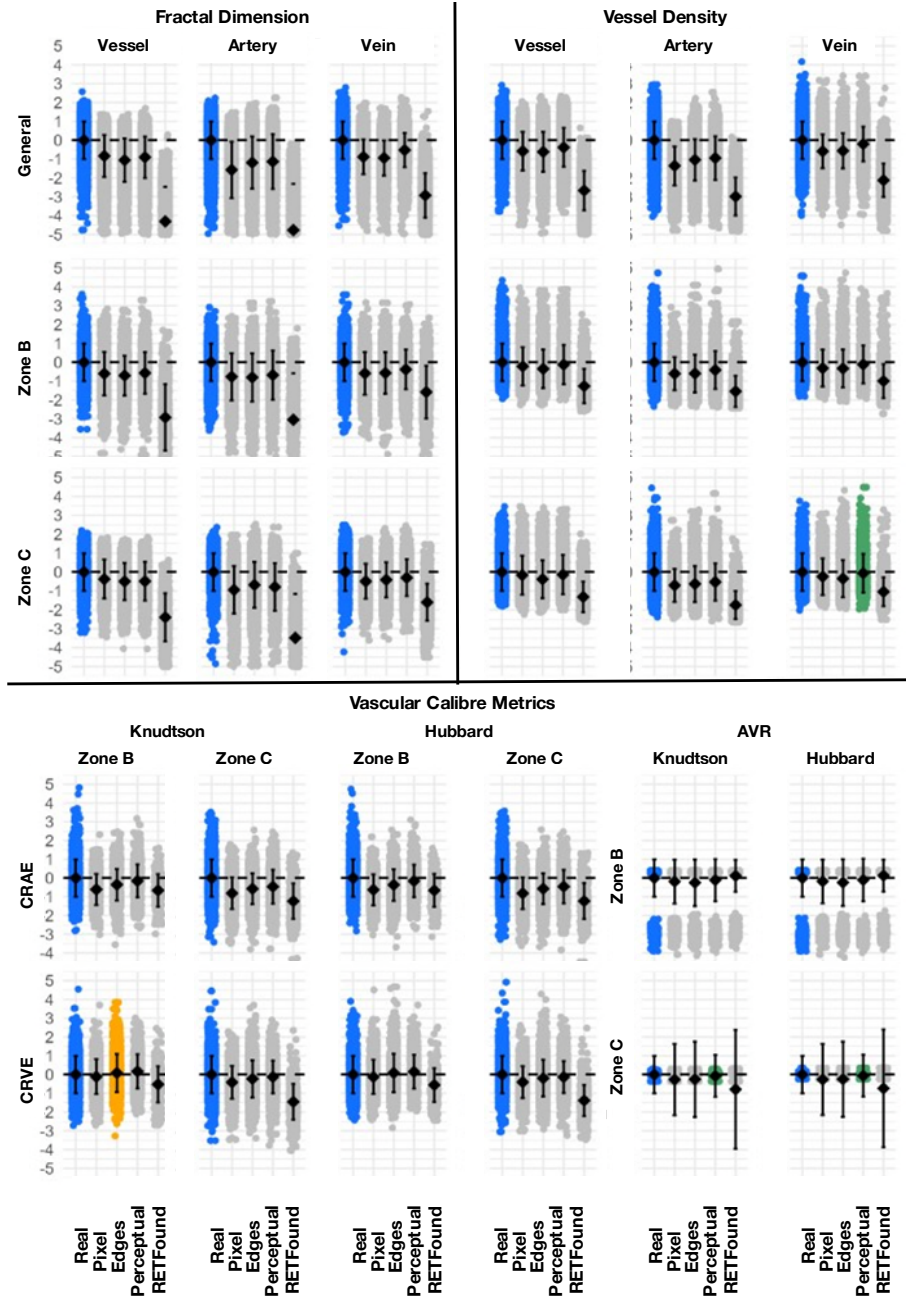

**Fig. 6.** Comparison of retinal features related to fractal dimension, vessel density and vascular calibre metrics. Each boxplot shows the mean and standard deviation for each model. Non-grey distributions indicate samples' overlap with the real data distribution, as supported by a permutation test (p-value > 0.05). All features are z-score normalized relative to the real sample.

## 5  Discussion

This paper compares different colour fundus image generation strategies enforcing either perceptual similarity (Perceptual loss), vessel continuity (Edge loss) or domain specific information (RETFound loss). We tested the proposed models on a range of validation tasks, starting from the domain-invariant validation on fidelity metrics, and finishing by direct comparison of extracted retinal features. Our work suggests that leveraging deep features of foundational models trained on domain data **does not** yield samples, which are more realistic or better preserve morphological and clinical information within colour fundus images. Conversely, models which are supervised by a perceptual loss, i.e., by models trained on natural imaging data, better learn to generate images that are both more realistic and morphologically and clinically correct. This is a surprising result, given the demonstrated use of RETFound encoder features for risk prediction of various cardiovascular events, e.g., myocardial infarction or ischaemic stroke [3]. Whilst the RETFound loss consistently underperformed, the autoencoder supervised by a simple edge detection filter achieved results comparable to the autoencoder supervised by deep features from the VGG-16 network - which serves as the backbone for the perceptual loss and was trained on over a million images from ImageNet. This outcome suggests that large pretrained networks, extensive datasets, or high computational resources may not always be required to achieve competitive performance.

Nevertheless, even the best-performing autoencoder in the study, VQ-GAN with perceptual loss, produced samples that underperformed relative to real samples in downstream tasks based on extracted retinal features. This issue may stem from the low-resolution regime used in our study. Typically, colour fundus imaging is performed at high resolution, and estimating vascular thickness from lower resolution data can inherently lead to inaccuracies. The underperformance of RETFound compared to VGG-16 may also stem from differences in model architecture, with RETFound being a vision transformer and VGG-16 a CNN. Notably, however, perceptual losses based on vision transformers have been shown to outperform their pixel-based loss counterparts [33]. Another possible reason for the relative drop in RETFound loss performance compared to perceptual loss is the difference in the number of activation layers used, with RETFound loss relying only on the final activation layer. Nevertheless, future research should explore alternative strategies for colour fundus synthetic sample generation that better preserve topological structures and retinal biomarkers than perceptual-based loss functions, such as those based on VGG-16 or RETFound. Furthermore, it would be beneficial to explore further whether CNN-based models are inherently more effective at preserving deep imaging features, which could be leveraged as imaging biomarkers in downstream tasks or serve as a backbone for a perceptual-style loss function.

## 6      Acknowledgments